\documentclass[12pt,a4paper]{article}
\usepackage{t1enc}
\usepackage[dvips,final]{graphicx} 
\usepackage{tabularx} 
\usepackage{amsfonts}
\usepackage{amssymb,amsmath}

\def\du{\unskip\smash{\lower 1.4ex \hbox{\char34}}\kern-.2ex}
\def\hu{\kern-.2ex\hbox{\char92}}

\def\XXint#1#2#3{{\setbox0=\hbox{$#1{#2#3}{\int}$}
     \vcenter{\hbox{$#2#3$}}\kern-.5\wd0}}

\newcommand{\bdis}{\begin{displaymath}}
\newcommand{\edis}{\end{displaymath}}
\newcommand{\be}{\begin{equation}}
\newcommand{\ee}{\end{equation}}

\newtheorem{pr*}[thm]{*}

\begin{document}
\baselineskip=6mm
\newpage

\title{Motion of a testing particle in gravitational field of a ring}
\author{Michal Demetrian\footnote{email: {\tt{demetrian@fmph.uniba.sk}}} \\
Comenius University
Mlynska Dolina M 105 Bratislava 4 \\
842 48 Slovak Republic}
\maketitle

\abstract{The planar motion of a testing body in the field of gravity of the massive ring is studied. The "perihelion"
precession due to non-relativistic reasons is demonstrated on this simple example of planar motion in the radially symmetric
(in the plane of motion) situation. The ring as the source of the
gravitational field can serve as the toy model for quasi-planar sources of gravity appearing in the
galactic dynamics, etc. The results are used to
estimate the real perihelion shift of the planet Mercury due to gravitational quadrupole moment of the Sun.}

\section{The potential}

Let us consider the thin (one dimensional) homogenous massive ring of radius $R$ with total mass $M$. We consider the
cartesian coordinates $x,y,z$ such that the plane $Oxy$ coincides with the plane defined by the ring itself, and the
origin of our coordinates coincides with the center of the ring. Our first task is to compute the Newtonian potential
$V=V(x,y)$ of the gravitational field of the ring in the plane $Oxy$. Because of the rotational invariance of the problem,
the potential is the function of radial distance $r=\sqrt{x^2+y^2}$ only. According to the Newton's gravitational law we
have
\begin{align*}
&V(r)=-\frac{\kappa M}{2\pi R}\int_0^{2\pi}\frac{R{\rm d}\psi}
{\left[(R\cos(\psi)-x)^2+(R\sin(\psi)-y)^2\right]^{1/2}}=
\left|\begin{array}{l} x=r\cos(\phi) \\ y=r\sin(\phi)\end{array}\right|=& \\
&-\frac{\kappa M}{2\pi}\int_0^{2\pi}\frac{{\rm d}\psi}
{\left[ R^2+r^2-2Rr\cos(\psi-\phi)\right]^{1/2}}=
-\frac{\kappa M}{2\pi}\int_0^{2\pi}\frac{{\rm d}\psi}
{\left[ R^2+r^2-2Rr\cos(\psi)\right]^{1/2}}=& \\
&-\frac{\kappa M}{2\pi}\frac{1}{\sqrt{R^2+r^2}}\int_0^{2\pi}\frac{{\rm d}\psi}
{\left[ 1-\frac{2Rr}{R^2+r^2}\cos(\psi)\right]^{1/2}} . &
\end{align*}
We see that the potential has the singularity at $r=R$ (at the ring). It holds:
\bdis
X\equiv \frac{2Rr}{R^2+r^2}=2\frac{\frac{r}{R}}{1+\left(\frac{r}{R}\right)^2}\leq 1 ,
\edis
and the sign of equality takes place only for $r=R$. We remind the definition of the complete elliptic integral of
the first kind
\bdis
K(x)=\int_0^{\pi/2}\frac{{\rm d}y}{\left[ 1-x\sin^2(y)\right]^{1/2}}, \quad 0\leq x<1 ,
\edis
with help of which we can express our potential as follows
\begin{align*}
&\int_0^{2\pi}\frac{{\rm d}\psi}
{\left[ 1-X\cos(\psi)\right]^{1/2}}=2\int_0^{\pi}\frac{{\rm d}\psi}{\left[ 1-X\cos(\psi)\right]^{1/2}}=
2\int_0^{\pi}\frac{{\rm d}\psi}{\left[ 1+X\cos(\psi)\right]^{1/2}}=& \\
&2\int_0^{\pi}\frac{{\rm d}\psi}{\left[ 1+X\left(\cos^2(\psi/2)-\sin^2(\psi/2)\right)\right]^{1/2}}=
2\int_0^{\pi}\frac{{\rm d}\psi}{\left[ 1+X-2X\sin^2(\psi/2)\right]^{1/2}}=& \\
&\frac{4}{\sqrt{1+X}}\int_0^{\pi/2}\frac{{\rm d}\xi}{\left[1-\frac{2X}{1+X}\sin^2(\xi)\right]^{1/2}}=
\frac{4}{\sqrt{1+X}}K\left(\frac{2X}{1+X}\right) . &
\end{align*}
Finally, we obtain the potential of the ring in the form
\be \label{potential}
V(r)=-\frac{2\kappa M}{\pi R}\frac{1}{1+\frac{r}{R}}K\left(\frac{4\frac{r}{R}}{\left( 1+\frac{r}{R}\right)^2}\right) .
\ee
Having the potential we can derive the gravitational force that acts upon a testing body of mass $m$:
\be \label{force}
f(r)\equiv m g(r)=-m V'(r)=-\frac{\kappa m M}{\pi R^2}
\frac{\left( 1+\frac{r}{R}\right)E\left(\frac{4\frac{r}{R}}{\left( 1+\frac{r}{R}\right)^2}\right)
+\left( 1-\frac{r}{R}\right)K\left(\frac{4\frac{r}{R}}{\left( 1+\frac{r}{R}\right)^2}\right)}
{\frac{r}{R}\left[\left(\frac{r}{R}\right)^2-1\right]} ,
\ee
where $E$ is the complete elliptic integral defined for $0\leq x\leq 1$ by
\bdis
E(x)=\int_0^{\pi/2}\sqrt{1-x\sin^2(y)}{\rm d}y
\edis
and we have denoted by $g$ the gravitational acceleration.

\begin{figure}[h]
\centering
\includegraphics[]{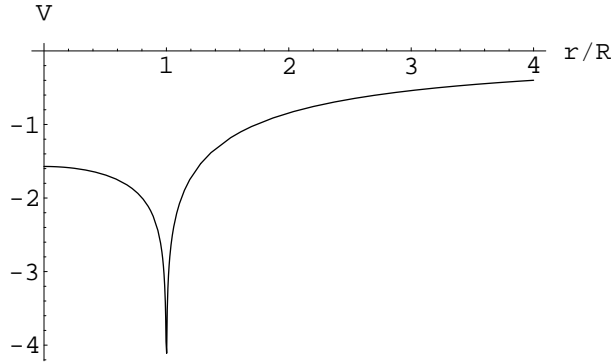}
\caption{Plot of the potential (\ref{potential}) as a function of the fraction $r/R$ with $2\kappa M/(\pi R)=1$. }
\label{potfig}
\end{figure}

For the next purposes we will write down the approximative expressions of the potential in the following three
cases:
\begin{itemize}
\item $r/R\gg 1$:
\be \label{potlim1}
V(r)=-\frac{\kappa M}{r}\left\{ 1+\frac{1}{4}\left(\frac{R}{r}\right)^2+
\frac{9}{64}\left(\frac{R}{r}\right)^4+\mathcal{O}\left(\frac{R}{r}\right)^6\right\}
\ee
\item $r/R\ll 1$:
\be \label{potlim2}
V(r)=-\frac{\kappa M}{R}\left\{ 1+\frac{1}{4}\left(\frac{r}{R}\right)^2+
\frac{9}{64}\left(\frac{r}{R}\right)^4+\mathcal{O}\left(\frac{r}{R}\right)^6\right\}
\ee
\item $r/R\approx 1$:
\begin{eqnarray} \label{potlim3}
& &
V(r)=-\frac{2\kappa M}{\pi R}
\left\{
\left[\frac{\ln(2)-\gamma_E}{2}-\frac{1}{2}\frac{\Gamma'(1/2)}{\Gamma(1/2)}-\frac{1}{2}\ln\left|\frac{r}{R}-1\right|\right]
+\right. \nonumber \\
& &
\left. \left[\frac{1+\gamma_E-\ln(2)+\Gamma'(1/2)/\Gamma(1/2)}{4}+\frac{1}{4}\ln\left|\frac{r}{R}-1\right|\right]
\left(\frac{r}{R}-1\right)+\dots\right\} ,
\end{eqnarray}
where $\gamma_E\approx 0.577216$ is the Euler's $\gamma$ constant and $\Gamma$ is the Euler's Gamma function.
\end{itemize}
Equation (\ref{potlim1}) tells us that at large distances from the ring the field is almost identical with the
field of point mass $M$ and the leading correction term (proportional to $(r/R)^2$) is the projection of the
quadrupole field into the plane of the ring.
Equation (\ref{potlim2}) shows that the center of the ring is an unstable equilibrium. If a testing particle is displaced
a little bit from the center then it is forced to move towards the ring, and the force is proportional to the
distance from the center (inverted linear harmonic potential). Finally, equation (\ref{potlim3}) shows the feature
of the potential close the ring. Namely, there is the logarithmic divergency of the potential as $r\to R$.

\section{Equations of motion and the conservation laws}

An easy way how to obtain equations of motion is to write down the Lagrange function (the Lagrangian) of
the testing particle with the mass $m$ in polar coordinates $(r,\phi)$ (we will consider the motion in the plane of the
ring only, as we have already mentioned above):
\be \label{lagrangian}
L=\frac{1}{2}m\left( \dot{r}^2+r^2\dot{\phi}^2\right)-m V(r) ,
\ee
where the dot stands for the time derivative, and derive the equations of motion as the Euler - Lagrange equations
\be \label{eqsmotion}
\ddot{r}+V'(r)=0, \quad r^2\ddot{\phi}+2r\dot{r}\dot{\phi}=0 .
\ee
The order of this system of two differential equations can be reduced using the conservation laws.
First of all, the energy $E$ is conserved:
\be \label{econserv}
\frac{1}{2}m\left( \dot{r}^2+r^2\dot{\phi}^2\right)+m V(r)=E.
\ee
The angular momentum vector must have the form $(0,0,\mathcal{L})$. Since $L$ does not depend on $\phi$,
we have the following expression for the $z$- component of the angular momentum:
\be \label{lconserv}
mr^2\dot{\phi}=\mathcal{L} .
\ee
The previous two equations can be combined into one expressing the radial velocity $\dot{r}$ as the
function of the radial position $r$ and the conserved quantities $E$ and $\mathcal{L}$:
\be \label{rdoteq}
\left|\dot{r}\right|=\left[\frac{2E}{m}-2V(r)-\frac{\mathcal{L}^2}{m^2r^2}\right]^{1/2} .
\ee
The inequality
\bdis
\frac{2E}{m}-2V(r)-\frac{\mathcal{L}^2}{m^2r^2}\geq 0
\edis
defines, for given values of $E$ and $\mathcal{L}$, classically accessible region, i.e. the minimal
$r_m$ and the maximal $r_M$ value of radial coordinate of considered particle.

\section{Radial motion (free fall)}

\begin{figure}[h]
\centering
\includegraphics[]{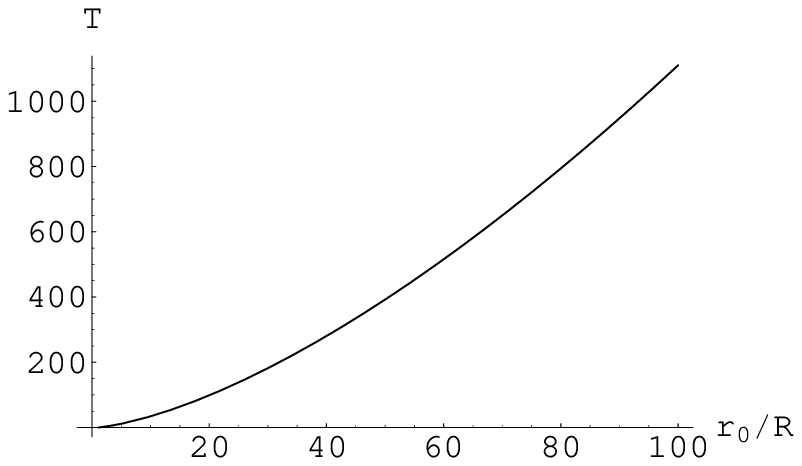}
\includegraphics[]{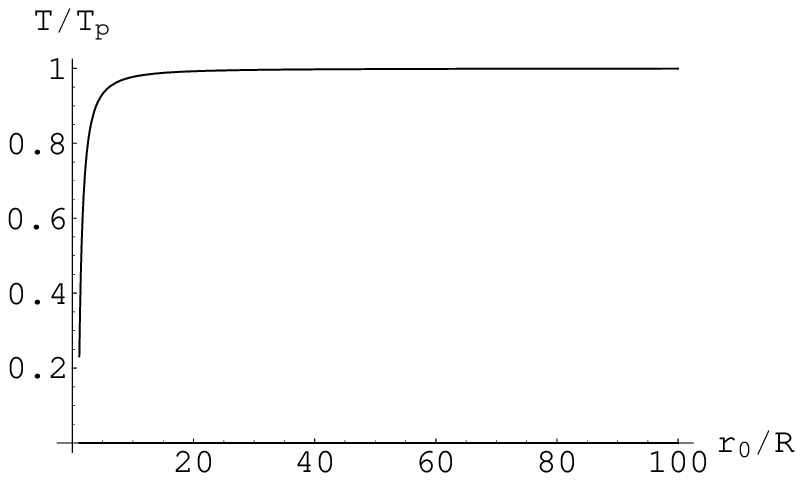}
\caption{Left graph shows the dependence of the free fall duration $T$
on initial distance $r_0/R$ according (\ref{Tfall}), where wee have
chosen for definiteness: $\kappa M=1$ and $R=1$. Right graph shows the fraction $T/T_p$ with the parameters
chosen in the same way as in the left graph. We see that the quantity $T$ approaches relatively $T_p$ as
$r_0/R$ grows. More detailed analysis would show that the difference $T_p-T$ approaches a positive constant of the
order of unity as $r_0/R$ runs to infinity.}
\label{tfig}
\end{figure}

The radial motion or the free fall is characterized by the condition $\mathcal{L}=0$. We will consider the
situation when the initial (radial) velocity equals zero. This means that $E/m=V(r_0)$, where $r_0$ is the
initial position, and equation (\ref{rdoteq}) becomes
\bdis
\left|\dot{r}\right|=\sqrt{2}\left[ V(r_0)-V(r)\right]^{1/2} .
\edis
Our task is to compute the time of the fall to the ring $T$. First, we will consider the case when $r_0>R$. Thus,
\bdis
\dot{r}=-\sqrt{2}\left[ V(r_0)-V(r)\right]^{1/2} \Rightarrow
T(r_0)=\frac{1}{\sqrt{2}}\int_R^{r_0}\frac{{\rm d}r}{\left[ V(r_0)-V(r)\right]} .
\edis
Explicitly,
\begin{eqnarray} \label{Tfall}
& &
T(r_0)=\frac{1}{2}\left[\frac{\pi R^3\left( 1+\frac{r_0}{R}\right)}
{\kappa M K\left(\frac{4\frac{r_0}{R}}{(1+\frac{r_0}{R})^2}\right)}\right]^{1/2}
\int_1^{r_0/R}\frac{{\rm d}X}
{\left[\frac{1+\frac{r_0}{R}}{1+X}
\frac{K\left(\frac{4X}{(1+X)^2}\right)}{K\left(\frac{4r_0/R}{(1+r_0/R)^2}\right)}-1\right]^{1/2}} .
\end{eqnarray}
If the initial position of the particle is inside the ring ($r_0<R$) then
\bdis
\dot{r}=+\sqrt{2}\left[ V(r_0)-V(r)\right]^{1/2} \Rightarrow
T(r_0)=\frac{1}{\sqrt{2}}\int_R^{r_0}\frac{{\rm d}r}{\left[ V(r_0)-V(r)\right]} ,
\edis
thus (\ref{Tfall}) holds also in this case. The free fall at the ring from outside can be compared with the
radial fall at the point mass $M$ placed in the center of the ring. The time of duration of such a fall is
given by
\be \label{tpfall}
T_p(r_0)=\int_0^{r_0}\frac{{\rm d}r}{\left[2\kappa M\left(\frac{1}{r}-\frac{1}{r_0}\right)\right]^{1/2}}=
\left[\frac{r_0^3}{2\kappa M}\right]^{1/2}\int_0^1\sqrt{\frac{\xi}{1-\xi}}{\rm d}\xi=
\frac{\pi}{2}\left[\frac{r_0^3}{2\kappa M}\right]^{1/2} .
\ee
The quantities $T$ and $T_p$ are compared in the fig. \ref{tfig}.

\section{Circular orbits}

For a testing body on a circular orbit the centripetal force $f_c$ must be equal to the gravitational force (\ref{force}).
This means that
\bdis
f_c=ma_c=-m\dot{\phi}^2r=-\frac{\mathcal{L}^2}{mr^3} \Rightarrow
-\frac{\mathcal{L}^2}{mr^3}=g .
\edis
The last equation can obviously be fulfilled for any $r>R$ but, of course, cannot be fulfilled for any $r<R$, i.e. the testing
particle can orbit the ring along any circle from outside.

\section{Small deformations of circular orbits - "perihelion"\ shift}

\begin{figure}[h]
\centering
\includegraphics[]{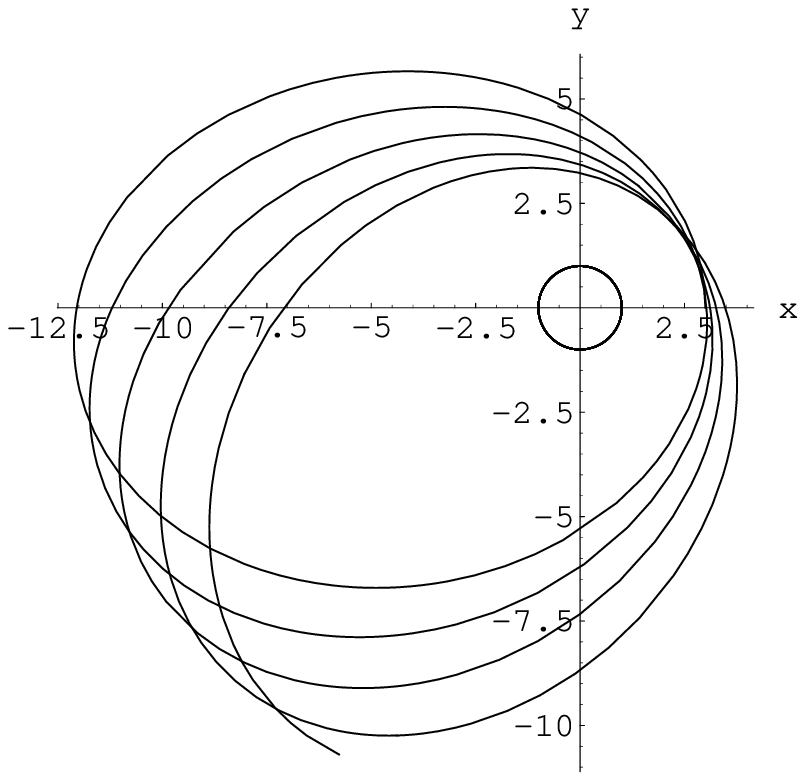}
\includegraphics[]{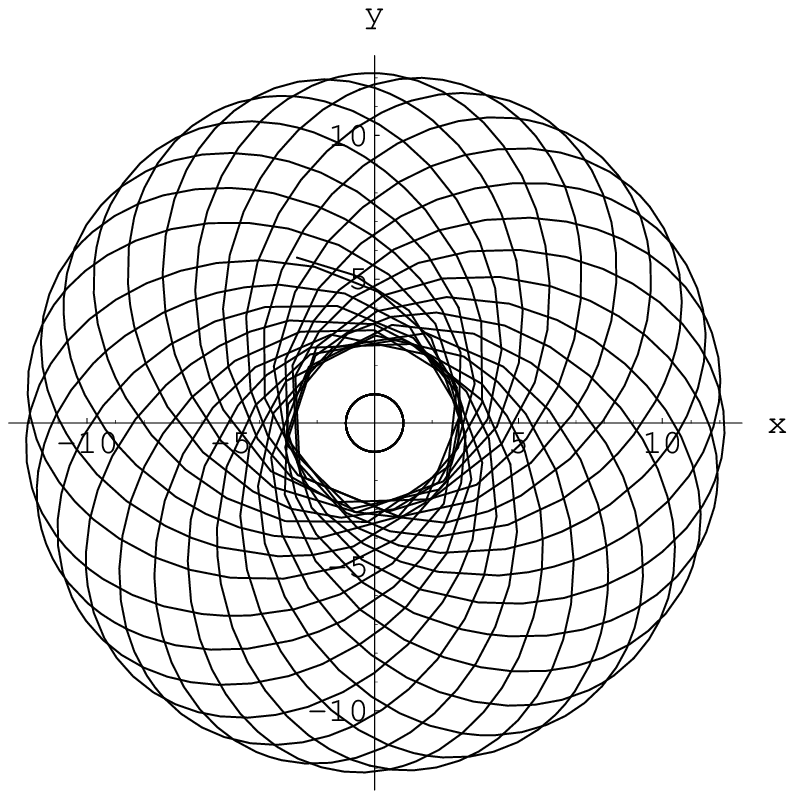}
\caption{Plot of the trajectory of a testing particle in the field of massive ring. The radius of the ring is $R=1$ and
$E$ and $\mathcal{L}$ were chosen so that the related Keplerian trajectory would be described by
$r_m=3$ and $r_M=15$. The body of the ring is depicted as the central circle. The particle orbits the ring
anti-clockwise, we see, from the left figure, that the "perihelion" moves in the same direction. }
\label{shiftfig}
\end{figure}

In this section we will consider the close-to-circular trajectory of a testing particle sufficiently distanced from the
ring. This means that the potential of the ring can be approximated by the first two terms of (\ref{potlim1}):
\be \label{vapprox}
V\approx -\frac{\kappa M}{r}\left( 1+\frac{1}{4}\frac{R^2}{r^2}\right)\equiv -\frac{\kappa M}{r}+\frac{\gamma}{r^3},
\quad \gamma=-\frac{1}{4}\kappa MR^2 .
\ee
It is well-known that the closed non-circular trajectories of testing particle in a central field exist only in two
cases: $V\sim r^2$ and $V\sim -1/r$, see eg. \cite{ll} or detailed proof in \cite{arnold}.
Therefore our close-to-circular orbit will not be closed, however for a small
deviation from the circle it can be regarded as slowly rotating ellipse. We are interested in the angle connecting
the two nearest positions of the particle and the ring
(the point of closest approach of a planet (from our solar system, of course) and the Sun is called perihelion)
and the center of our coordinate system. \\
Combining the eqs. (\ref{lconserv}) and (\ref{rdoteq}) we easily can derive that the change of the polar angle
$\Delta \phi$ between the two forthcoming closest approaches of the particle with respect to the ring is given by
\begin{align} \label{dphi1}
&\Delta\phi=2\int_{r_m}^{r_M}
\frac{\frac{\mathcal{L}}{m}\frac{1}{r^2}{\rm d}r}{\sqrt{\frac{2E}{m}-2V-\frac{\mathcal{L}^2}{m^2r^2}}}=
2\mathcal{L}\int_{r_m}^{r_M}\frac{\frac{1}{r^2}{\rm d}r}{\sqrt{2m(E-V)-\frac{\mathcal{L}^2}{r^2}}}=& \nonumber \\
&-2\frac{\partial}{\partial\mathcal{L}}\int_{r_m}^{r_M}\sqrt{2m(E-V)-\frac{\mathcal{L}^2}{r^2}}{\rm d}r , &
\end{align}
where $r_m$ and $r_M$ are the minimal and the maximal values of the radius vector of the particle, respectively. The
angle difference $\Delta \phi$ can be written as: $\Delta\phi=2\pi+\delta\phi$, where the difference $2\pi$ comes
from the Newton potential and additional term $\delta\phi$ is called perihelion shift in the mechanics
of planetary motion around the Sun or the periastron shift in the mechanics of binary stars. An illustration of the
"perihelion shift" in our case is given in fig. \ref{shiftfig}.

Using (\ref{vapprox}) and
the Taylor expansion of the subintegral expression in (\ref{dphi1}) we obtain
\begin{align} \label{dphi2}
&\delta\phi=\frac{\partial}{\partial\mathcal{L}}\frac{2m}{\mathcal{L}}\int_0^\pi r^2\frac{m\gamma}{r^3}{\rm d}\phi=
2m^2\gamma\frac{\partial}{\partial\mathcal{L}}\frac{1}{\mathcal{L}}\int_0^\pi \frac{1}{r(\phi)}{\rm d}\phi , &
\end{align}
where $r=r(\phi)$ is the elliptic trajectory in the Newtonian field of a point mass. For this trajectory we have
\begin{align*}
&r=\frac{p}{1+\epsilon\cos(\phi)}& & p=\frac{\mathcal{L}^2}{\kappa m^2M}&
&\epsilon=\left[ 1+\frac{2E\mathcal{L}^2}{\kappa^2m^3M^2}\right]^{1/2}, &
\end{align*}
where $\epsilon$ is the eccentricity and $p$ is the parameter of the ellipse with the semi-axis
\begin{align*}
&a=\frac{p}{1-\epsilon^2}& &b=\frac{p}{\sqrt{1-\epsilon^2}} , &
\end{align*}
and
\begin{align*}
&r_m=\frac{p}{1+\epsilon}& &r_M=\frac{p}{1-\epsilon} . &
\end{align*}
Finally, we have
\begin{align} \label{dphi3}
&\delta\phi=2\pi m^2\gamma\frac{\partial}{\partial\mathcal{L}}\frac{1}{\mathcal{L}p}=
-6\pi m^2\gamma\frac{1}{\mathcal{L}^2p}=-6\pi\frac{(\kappa mM)m^3\gamma}{\mathcal{L}^4}=
\frac{3}{2}\pi\frac{\kappa^2m^4M^2R^2}{\mathcal{L}^4}
, &
\end{align}
or using the relation
\bdis
a(1-\epsilon^2)=\frac{\mathcal{L}^2}{\kappa m^2M}
\edis
we can express the perihelion shift (\ref{dphi3}) with help of geometrical parameters of considered perturbed ellipse:
\bdis
\delta\phi=\frac{3}{2}\pi\frac{R^2}{a^2(1-\epsilon^2)^2} .
\edis

\section{Discussion}

We have studied in details the motion of a testing particle in the gravitational field of the massive planar ring.
As the main aim, we have determined the shift of the perihelion of the testing particle with
non-circular orbit
with sufficiently large radius. For $r/R\gg 1$ the gravitational field of the ring is practically identical
with the field of central monopole of the mass $M$ and the quadrupole field, see (\ref{potlim1}).
The formula (\ref{dphi3})
describes, in fact, the perihelion shift for the particle in the planar motion due
to quadrupole perturbation of Newton potential. Hence, it can be uses to estimate the perihelion shift of the planet
Mercury due to Solar quadrupole moment, considering the approximation that
the trajectory of Mercury is placed in the plane of the solar
equator. Following the results about the multipole expansion of the gravitational field of the Sun we have
\bdis
V_\odot=-\frac{\kappa M_\odot}{r}\left\{ 1-J_2\frac{R_\odot^2}{r^2}P_2(\cos(\vartheta))+
\mathcal{O}\left(R^3_\odot/r^3\right)\right\} ,
\edis
where $M_\odot$ is the mass of the Sun, $R_\odot$ is the mean radius of the Sun,
$P_2(z)=-1/2+3/2z^2$ is the Legendre polynomial that depends on the azimutal angle $\vartheta$ and
$J_2=(2.18\pm 0.06)\times 10^{-7}$ is the dimensionless quadrupole moment of the Sun - for a long time the value of
solar gravitational quadrupole was very uncertain, this value is from \cite{stix}.
In the equator of the Sun plane
we have $\vartheta=\pi/2$ and therefore in the mentioned plane the potential has the form
\bdis
V_\odot=-\frac{\kappa M_\odot}{r}\left\{ 1+\frac{J_2}{2}\frac{R_\odot^2}{r^2}+
\mathcal{O}\left(R^3_\odot/r^3\right)\right\} .
\edis
So, we can use the formula (\ref{dphi3}) to compute the Mercury's perihelion shift due to the Sun's quadrupole moment
identifying the constant $\gamma$ as follows:
\be \label{mercury}
\gamma=-\frac{1}{2}J_2\kappa M_\odot R^2_\odot .
\ee
Finally, taking into account the Mercury's parameters: the mass $m\approx 3.302\times 10^{23}kg$,
distance from the Sun in the
perihelion $r_m\approx4.6\times 10^7km$, orbital velocity in the perihelion $v_M\approx 59 km/s$
($\mathcal{L}=mr_mv_M$, orbital period is about $88$ days),
and the mass of the Sun: $M_\odot\approx 2\times 10^{30}kg$ and its radius $R_\odot\approx 7\times 10^5 km$, we have
\be \label{mercury1}
\delta\phi=3\pi J_2\frac{\kappa^2 m^4M_\odot^2R_\odot^2}{\mathcal{L}^2}=
3\pi J_2\frac{\kappa^2 M_\odot^2R_\odot^2}{r_m^4v_M^4}\approx 3.3\times 10^{-10} .
\ee
This means that during the period of $100$ years, the perihelion of the Mercury is shifted in the direction of
planetary rotation around the Sun
in approximately $0.28$ arcsec. This value can be compared with the well-known value of the
relativistic shift of the perihelion that is about $35$ arcsec per a century. Our value (\ref{mercury1}) is
quite less that is in accordance with accepted fact that the Solar quadrupole does not effect
essentially the Mercury's orbit in the time interval about $100$ years.

\end{document}